**TiO$_2$ Nanocrystals Grown on Graphene as Advanced Photocatalytic Hybrid Materials**

Yongye Liang[†], Hailiang Wang[†], Hernan Sanchez Casalongue, Zhuo Chen and Hongjie Dai*

Department of Chemistry and Laboratory for Advanced Materials, Stanford University, Stanford, California 94305, USA

Page Numbers. The font is ArialMT 16 (automatically inserted by the publisher)

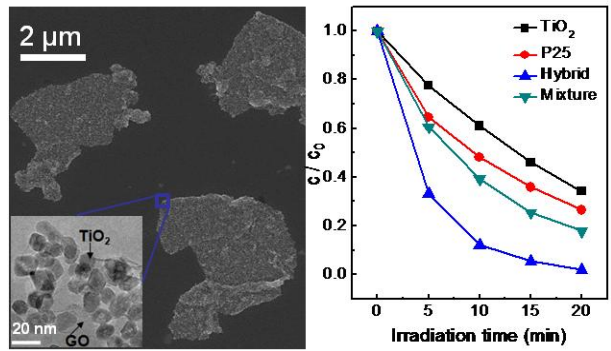

Graphene/TiO$_2$ nanocrystals hybrid is successfully prepared by directly growing TiO$_2$ nanocrystals on graphene oxide (GO) sheets by a two-step method, in which TiO$_2$ was coated on GO sheets by hydrolysis first and crystallized into anatase nanocrystals by hydrothermal treatment in second step. The prepared graphene/TiO$_2$ nanocrystals hybrid has demonstrated superior photocatalytic activity in degradation of rhodamine B over other TiO$_2$ materials, showing an impressive 3-fold photocatalytic enhancement over P25.

# TiO$_2$ Nanocrystals Grown on Graphene as Advanced Photocatalytic Hybrid Materials

Yongye Liang[†], Hailiang Wang[†], Hernan Sanchez Casalongue, Zhuo Chen and Hongjie Dai(✉)

Department of Chemistry and Laboratory for Advanced Materials, Stanford University, Stanford, California 94305, USA
[†] These authors contributed equally.

## ABSTRACT

Graphene/TiO$_2$ nanocrystals hybrid is successfully prepared by directly growing TiO$_2$ nanocrystals on graphene oxide (GO) sheets. The direct growth of nanocrystals on GO sheets was achieved by a two-step method, in which TiO$_2$ was coated on GO sheets by hydrolysis first and crystallized into anatase nanocrystals by hydrothermal treatment in second step. Slow hydrolysis reaction through the use of EtOH/H$_2$O mixed solvents and addition of H$_2$SO$_4$ allows the selectively growing TiO$_2$ on GO and suppressing free growth in solution. The method offers easy access to the GO/TiO$_2$ nanocrystals hybrid with well controlled coating and strong interactions between TiO$_2$ and the underlying GO sheets. The strong coupling could lead to advanced hybrid materials for various applications including photocatalysis. The prepared graphene/TiO$_2$ nanocrystals hybrid has demonstrated superior photocatalytic activity in degradation of rhodamine B over other TiO$_2$ materials, showing an impressive 3-fold photocatalytic enhancement over P25. It is expected that the hybrid material could also be promising for various other applications including lithium ion battery where strong electrical coupling to TiO$_2$ nanoparticles is essential.

The interesting electrical and mechanical properties, and high surface area make graphene a novel substrate for forming hybrid structures with various nanomaterials [1-3]. Graphene hybrids with metal oxides, metals and polymers have been developed recently for various applications [4-6]. Nanocrystal growth on graphene sheet is an important approach to produce nano-hybrids since controlled nucleation and growth affords optimal chemical interactions and bonding between nanocrystals and graphene sheets, leading to the strongest electrical and mechanical coupling within the hybrid. Several methods have been proposed to form nanocrystals on graphene sheets, such as electrochemical deposition [7], sol-gel process [8] and gas phase deposition [9, 10]. Recently, we developed a controlled, two-step solution phase synthesis of nanocrystals of Ni, Fe and Co hydroxides or oxides selectively on graphene sheets (no over growth in free solution) [1]. We demonstrated excellent performance of nickel hydroxide nanoplates grown on graphene for electrochemical pseudocapacitive energy storage utilizing the high electrical conductivity of graphene [4].

Here we present a two-step, direct synthesis of $TiO_2$ nanocrystals on graphene oxide (GO) and advanced photocatalytic properties of the resulting hybrid material. Due to the low cost, high stability and efficient photo-activity, $TiO_2$ has been widely used for photoelectrochemical and photocatalytic applications [11]. Several graphene/$TiO_2$ composites have been reported recently for lithium ion battery [12], photocatalysis [13] and dye sensitized solar cells [14], using either surfactant assisted growth or simple physical mixing of pre-synthesized $TiO_2$ particles and graphene. Here we achieve surfactant free growth of $TiO_2$ nanocrystals on graphene with optimal $TiO_2$-graphene coupling. By limiting the rate of hydrolysis, we obtain high degree of control to confine $TiO_2$ nanocrystal growth selectively on graphene oxide sheets, without any free growth of $TiO_2$ in solution. The resulting hybrid material showed superior photocatalytic degradation of rhodamine B and methylene blue over various other $TiO_2$ materials.

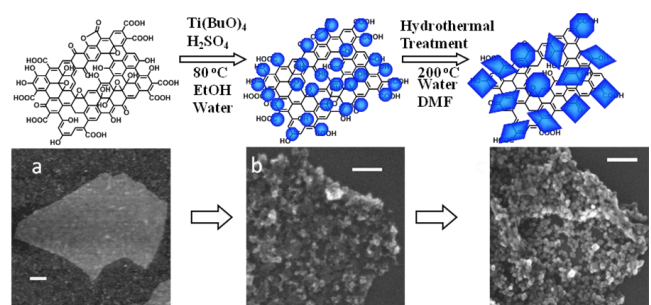

**Figure 1.** Synthesis of $TiO_2$ nanocrystals on GO sheets. Top panel: reaction schemes. (a) AFM image of a starting GO sheet. (b) SEM image of particles grown on a GO sheet after the first step hydrolysis reaction. (c) SEM image of $TiO_2$ nanocrystals on GO after hydrothermal treatment in the second step. The scale bars are 100 nm.

GO prepared by a modified Hummers method [15-18] was used as substrates for $TiO_2$ growth to form $TiO_2$/GO hybrid (containing ~10% GO by mass). The functional groups on GO provided reactive and anchoring sites for nucleation and growth of nanomaterials (Figure 1) [1]. In the first step reaction, fine particles of amorphous like $TiO_2$ (see XRD in Figure S-1 in Supporting Information) was coated on GO sheets by hydrolysis of $Ti(BuO)_4$ at 80 °C with the addition of $H_2SO_4$ in a EtOH/$H_2O$ (15/1, volume ratio) mixed solvent, designed to slow down the hydrolysis reaction. This led to selective growth of $TiO_2$ on GO (Figure 1b) without obvious free growth of $TiO_2$ particles in solution. Rapid hydrolysis occurred when only water was used as solvent or without the addition of $H_2SO_4$, in which $TiO_2$ particles grown in solution not associated with GO was observed (see Figure S-2). In the second step, we carried out a hydrothermal treatment of the amorphous $TiO_2$/GO at 200 °C in a mixed water/DMF solvent (Figure 1c), which led to crystallization of the coating material on GO into anatase nanocrystals (Figure 2b-d).

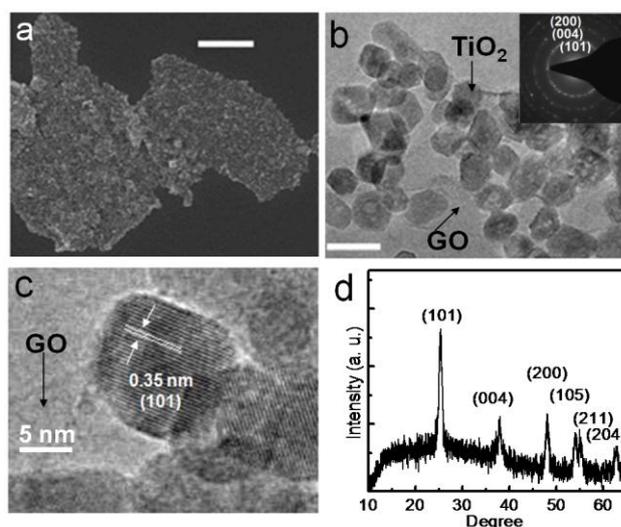

**Figure 2.** (a) SEM image, (b) low magnification and (c) high magnification TEM images of $TiO_2$ nanocrystals grown on graphene oxide sheets. The scale bar is 400 nm for the SEM image in (a) and 20 nm for the TEM image in (b). (d) An XRD spectrum of the Graphene/$TiO_2$ nanocrystal hybrid.

We used water/DMF (50/1) mixed solvent in the second step since DMF was found to facilitate dispersion of graphene sheets and reduce aggregation, suggested by the higher surface area (190 m$^2$/g) for the resulting hybrid than the material (134 m$^2$/g) obtained in pure water. Microscopy imaging (Figure1c, 2a&2b) revealed dense $TiO_2$ nanocrystals densely bound to GO sheets, without detachment even under sonication conditions. High resolution TEM showed that $TiO_2$ grown on GO is in a highly crystalline anatase phase, consistent with XRD (Figure 2c and 2d).

The density of $TiO_2$ nanocrystals coated on graphene was controlled by the feed ratio of $Ti(BuO)_4$ /GO, increasing as $Ti(BuO)_4$ amount increased (Figure 3). The average size of the $TiO_2$ nanocrystals depended on the EtOH/water ratio in the first step reaction, increased from ~15 nm in EtOH/Water (15/1) to ~30 nm in EtOH/Water (3/1) (Figure S-3) due to higher hydrolysis reaction rate when the water content was increased. This was accompanied by a decrease in surface area for $TiO_2$/GO synthesized in EtOH/water mixture with higher water concentrations (Table S-1).

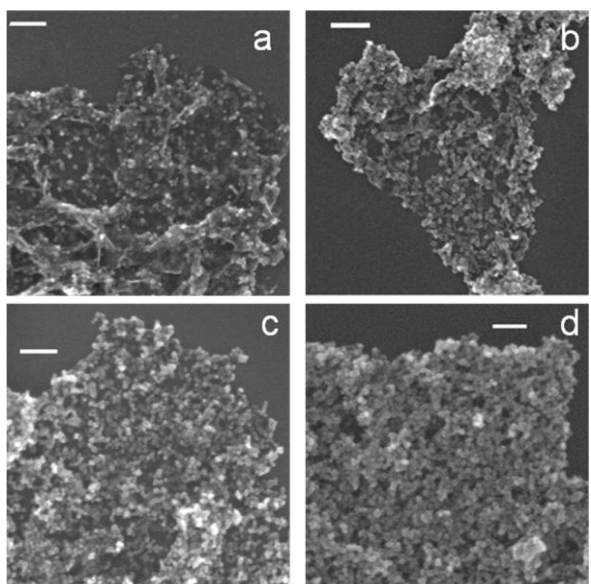

**Figure 3**. SEM images of $TiO_2$/GO hybrid made by various GO/$TiO_2$ mass ratios: (a) 1:1 (b) 1:3 (c) 1:9 (d) 1:18.. The scale bars are 100 nm. The $TiO_2$ coating density is controlled by varying the feed ratio of $Ti(BuO)_4$ /GO. The coating of $TiO_2$ nanocrystals on graphene is denser as $TiO_2$/GO feed ratio increases.

$TiO_2$ nanocrystals directly grown on graphene appeared to exhibit strong interactions with the underlying GO sheets and sustained sonication tests without dissociation from the sheets. The strong coupling may lead to advanced hybrid materials for various applications including photocatalysis. $TiO_2$ has been widely used as a photocatalytic semiconductor for organic decontamination. A limiting factor to the rate of photocatalysis is rapid electron-hole recombination [19]. To increase the charge separation lifetime, carbon nanotube (CNT)-$TiO_2$ hybrids have been produced to enhance photocatalysis rate through transfer of photo-excited electron to CNT [20].

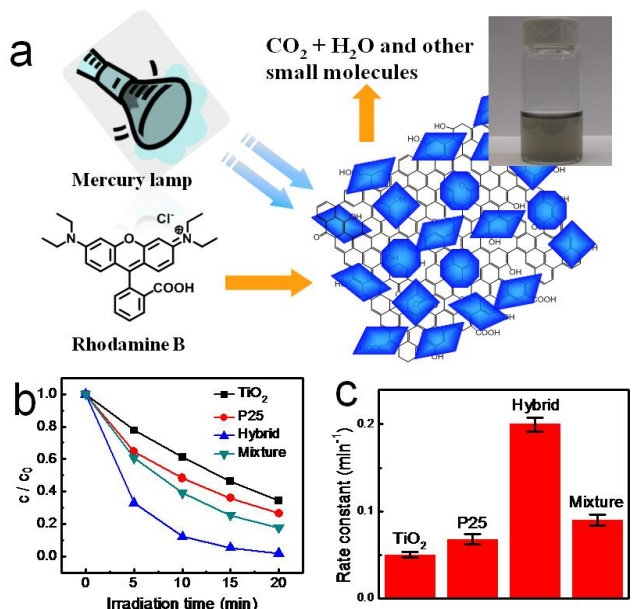

**Figure 4.** (a) Schematic illustration for the photo-degradation of rhodamine B molecules by the Graphene/$TiO_2$ nanocrystal hybrid under irradiation of mercury lamp. Inset shows the solution of Graphene/$TiO_2$ nanocrystal hybrid. (b) Photocatalytic degradation of rhodamine B monitored as normalized concentration change versus irradiation time in the presences of, synthesized TiO2, P25, Graphene/$TiO_2$ nanocrystal hybrid and Graphene/TiO2 mixture prepared from literature method. (c) Average reaction rate constant (min$^{-1}$) for the photodegradation of rhodamine B with synthesized TiO2, hybrid, P25 and Graphene/TiO2 mixture prepared from literature method. The error bars are based on more than three different batches of samples measured.

We tested the photocatalytic performance of our graphene/$TiO_2$ nanocrystals hybrid in photo-degradation of rhodamine B under UV irradiation (Figure 4a) and compared with free $TiO_2$ nanocrystals synthesized by the same method and P25 (a well-known commercial $TiO_2$ photocatalyst). The amount of $TiO_2$ content was kept the same for all these samples. The hybrid sample used for the photocatalytic measurement was prepared from EtOH/water = 15/1(volume ratio) in first step and water/DMF = 50/1 (volume ratio) in second step with $TiO_2$/GO = 9/1 (mass ratio). The degradation reaction was fit to a pseudo first-order reaction at low dye concentrations: $ln(c_0/c) = kt$, where $k$ is the apparent rate constant [21]. The average $k$ for the hybrid ($k$ = 0.200 $min^{-1}$) was found to be ~4 times of $TiO_2$ freely grown in solution ($k$ = 0.050 $min^{-1}$) and 3 times of P25 ($k$ = 0.068 $min^{-1}$). We also prepared simply mixed P25 and GO and hydrothermally treated the mixture as done in Ref. [13], and found the resulting mixture afforded $k$ = 0.090 $min^{-1}$, only half of our directly grown $TiO_2$/GO hybrid. (Figure 4b and 4c) The 3-fold photocatalytic enhancement of our hybrid over P25 was impressive, considering the low cost of GO and the highest enhancement factor of ~2 for CNT-$TiO_2$ composites over P25 [20]. The superior photocatalytic activity of our $TiO_2$/GO hybrid is also demonstrated in degradation of methylene blue. The average rate constant for the hybrid ($k$ = 0.128 $min^{-1}$) is much larger than those for P25 ($k$ =0.055 $min^{-1}$) and simple mixture of P25 and GO ($k$ =0.084 $min^{-1}$) (Figure S-4).

The high photocatalytic activity of our hybrid could be attributed to strong coupling between $TiO_2$ and GO to facilitate interfacial charge transfer (with GO as an electron acceptor) and prolong electron-hole recombination [22]. Also a contributing factor was the hybrid material exhibiting a higher BET surface area (190 $m^2$/g) than freely grown $TiO_2$ (121 $m^2$/g) and P25. The conjugated dye molecule could bind to large aromatic domains on GO sheet via π-π stacking, which could favor increased reactivity [19].

In conclusion, we have successfully grown $TiO_2$ nanocrystals on graphene with well-controlled coating density by a two step method. This method produced $TiO_2$/graphene hybrid with strong interactions between the two components. The resulting hybrid material shows superior photocatalytic activity. We expect that the hybrid material could also be useful for various other applications including $TiO_2$ based electrode for dye sensitized solar cells and lithium ion battery where strong electrical coupling to $TiO_2$ nanoparticles are also important. Our method could be further extended to grow other functional materials on graphene for advanced hybrid materials.

## Acknowledgements

This work is supported in part by Intel, MARCO-MSD and ONR

**Electronic Supplementary Material**: Supplementary material (Experimental details about making the graphene/$TiO_2$ hybrid and additional experimental data) is available in the online version of this article at http://dx.doi.org/10.1007/10.1007/s12274-***-****-* (automatically inserted by the publisher) and is accessible free of charge.

## References

[1] Wang, H. L.; Robinson, J. T.; Diankov, G.; Dai, H. J. Nanocrystal growth on graphene with various degrees of oxidation. *J. Am. Chem. Soc.* **2010**, *132*, 3270-3271.

[2] Si, Y. C.; Samulski, E. T. Exfoliated graphene separated by platinum nanoparticles. *Chem. Mater.* **2008**, *20*, 6792–6797.

[3] Lee, D. H.; Kim, J. E.; Han, T. H.; Hwang, J. W.; Jeon, S.; Choi, S. Y.; Hong, S. H.; Lee, W. J.; Ruoff, R. S.; Kim, S. O. Versatile carbon hybrid films composed of vertical carbon nanotubes grown on mechanically compliant graphene films. *Adv. Mater.* **2010**, *22*, 1247-1252.

[4] Wang, H. L.; Sanchez Casalongue, H.; Liang, Y. Y.; Dai, H. J. Ni(OH)$_2$ nanoplates grown on graphene as advanced electrochemical pseudocapacitor materials. *J. Am. Chem. Soc.* **2010**,*132*, 7472-7477.

[5] Yoo, E. J.; Okata, T.; Akita, T.; Kohyama, M.; Nakamura, J.; Honma, I. Enhanced electrocatalytic activity of Pt subnanoclusters on graphene nanosheet surface. *Nano Lett.* **2009**, *9*, 2255-2259.

[6] Murugan, A. V.; Muraliganth, T.; Manthiram, A. Rapid, facile microwave-solvothermal synthesis of graphene nanosheets and their polyaniline nanocomposites for energy storage. *Chem. Mater.* **2009**, *21*, 5004-5006.

[7] Williams, G.; Seger, B.; Kamat, P. V. $TiO_2$-graphene nanocomposites. UV-assisted photocatalytic reduction of graphene oxide. *ACS Nano* **2008**, *2*, 1487–1491.

[8] Wang, D. H.; Kou, R.; Choi, D. W.; Yang, Z. G.; Nie, Z. M.; Li, J.; Saraf, L. V.; Hu, D. H.; Zhang, J. G.; Graff, G. L.; Liu, J.; Pope, M. A.; Aksay, I. A. Ternary self-assembly

of ordered metal oxide−graphene nanocomposites for electrochemical energy storage. *ACS Nano*, **2010**, *4*, 1587-1595.

[9] Wang, X. R.; Tabakman, S. M.; Dai, H. J. Atomic layer deposition of metal oxides on pristine and functionalized graphene. *J. Am. Chem. Soc.*, **2008**, *130*, 8152-8153.

[10] Lu, G. H.; Mao, S.; Park, S.; Ruoff. R. S.; Chen, J. H. Facile, noncovalent decoration of graphene oxide sheets with nanocrystals. *Nano Res.* **2009**, *2*, 192-200.

[11] Chen, X. B.; Mao, S. S. Titanium dioxide nanomaterials: synthesis, properties, modifications, and applications. *Chem. Rev.* **2007**, *107*, 2891-2959.

[12] Wang, D. H.; Choi, D. W.; Li, J.; Yang, Z. G. ; Nie, Z. M.; Kou, R.; Hu, D. H.; Wang, C. M.; Saraf. L. V.; Zhang, J. G.; Aksay, I. A.; Liu, J. Self-assembled $TiO_2$–graphene hybrid nanostructures for enhanced Li-Ion insertion. *ACS Nano*, **2009**, *3*, 907-914.

[13] Zhang, H.; Lv, X. J.; Li, Y. M.; Wang, Y.; Li, J. H. P25-graphene composite as a high performance photocatalyst. *ACS Nano*, **2010**, *4*, 380-386.

[14] Yang, N. L.; Zhai, J.; Wang, D.; Chen, Y. S.; Jiang, L. Two-dimensional graphene bridges enhanced photoinduced charge transport in dye-sensitized solar cells. *ACS Nano*, **2010**, *4*, 887-894.

[15] Hummer, W. S.; Offeman, R. E. Preparation of graphitic oxide. *J. Am. Chem. Soc.* **1958**, *80*, 1339-1339.

[16] Wang, H. L.; Wang, X. R.; Li, X. L.; Dai, H. J. Chemical self-assembly of graphene sheets. *Nano Res.* **2009**, *2*, 336–342.

[17] Sun, X. M.; Liu, Z.; Welsher, K.; Robinson, J. T.; Goodwin, A.; Zaric, S.; Dai, H. J. Nano-graphene oxide for cellular imaging and drug delivery. *Nano Res.* **2008**, *1*, 203– 212.

[18] Wang, H. L.; Robinson, J. T.; Li, X. L.; Dai, H. J. Solvothermal reduction of chemically exfoliated graphene sheets. *J. Am. Chem. Soc.* **2009**, *131*, 9910-9911.

[19] Ravelli, D.; Dondi, D.; Fagnoni, M.; Albini, A. Photocatalysis. a multi-faceted concept for green chemistry. *Chem. Soc. Rev.* **2009**, *38*, 1999-2011.

[20] Woan, K.; Pyrgiotakis, G.; Sigmund, W. Photocatalytic carbon-nanotube-TiO2 composites. *Adv. Mater.* **2009**, *21*, 2233-2239.

[21] Wang, X. H.; Li, J. G.; Kamiyama, H.; Moriyoshi, Y.; Ishigaki, T. Wavelength-sensitive photocatalytic degradation of methyl orange in aqueous suspension over iron(III)-doped $TiO_2$ nanopowders under UV and visible light irradiation. *J. Phys. Chem. B*, **2006**, *110*, 6804-6809.

[22] Wang, W. D.; Serp, P.; Kalck, P.; Faria, J. L. Photocatalytic degradation of phenol on MWNT and titania composite catalysts prepared by a modified sol-gel method. *Appl. Catal. B* **2005**, *56*, 305-312.

———————————————

Address correspondence to Hongejie Dai, hdai@stanford.edu.